\documentclass[12pt]{iopart}
\usepackage{amssymb}
\usepackage{hyperref}
\usepackage{graphicx}
\usepackage{bm}
\usepackage[export]{adjustbox}
\usepackage{overpic}
\usepackage{ulem}
\usepackage{xcolor}

\begin{document}
\title{Many-body spin rotation by adiabatic passage in spin-1/2 XXZ chains of ultracold atoms}

\author{Ivana Dimitrova$^{1,*}$, Stuart Flannigan$^{2,**}$, Yoo Kyung Lee$^1$, Hanzhen Lin$^1$, Jesse Amato-Grill$^{1,***}$, Niklas Jepsen$^{1,***}$, Ieva \v{C}epait\.{e}$^2$, Andrew J. Daley$^2$ and Wolfgang Ketterle$^1$      }
\address{$^1$ Research Laboratory of Electronics, MIT-Harvard Center for Ultracold Atoms,
Department of Physics, Massachusetts Institute of Technology, Cambridge, Massachusetts 02139, USA.

$^2$ Department of Physics and SUPA, University of Strathclyde, Glasgow G4 0NG, United Kingdom.

$^*$ Present address: Department of Physics, Harvard University, Cambridge, Massachusetts 02138, USA.

$^{**}$ Present address: Strangeworks, Austin, Texas 78702, USA.

$^{***}$ Present address: QuEra Computing, Inc., Boston, MA 02135, USA.
}

\date{\today}

\begin{abstract}
Quantum many-body phases offer unique properties and emergent phenomena, making them an active area of research. A promising approach for their experimental realization in model systems is to adiabatically follow the ground state of a quantum Hamiltonian from a product state of isolated particles to one that is strongly-correlated. Such protocols are relevant also more broadly in coherent quantum annealing and adiabatic quantum computing. Here we explore one such protocol in a system of ultracold atoms in an optical lattice. A fully magnetized state is connected to a correlated zero-magnetization state (an xy-ferromagnet) by a many-body spin rotation, realized by sweeping the detuning and power of a microwave field. The efficiency is characterized by applying a reverse sweep with a variable relative phase. We restore up to $50\%$ of the original magnetization independent of the relative phase, evidence for the formation of correlations. The protocol is limited by the many-body gap of the final state, which is inversely proportional to system size, and technical noise. Our experimental and theoretical studies highlight the potential and challenges for adiabatic preparation protocols to prepare many-body eigenstates of spin Hamiltonians.

\end{abstract}


\maketitle

\section{Introduction}
The study of many-body quantum states is at the intersection of fundamental quantum physics and quantum technologies. Entangled and highly correlated quantum states lead to intriguing new properties of materials and are resources for quantum computation. A leading platform for engineering quantum spin Hamiltonians is provided by ultracold atoms in optical lattices \cite{bloch08}. Many recent studies in these systems have explored non-equilibrium quantum dynamics, often involving evolution from an initial state that is straight-forward to prepare on a single-particle level \cite{choi16, matt19, rispoli19, sanchez20, nature20, scherg21, wei22, oppong22}. The focus on such quench experiments reflects not only the strong general interest in such dynamics, but also the challenges of realising more complex many-body eigenstates. This is often related to the prevalence of low-lying excitations which lead to requirements of extremely low spin entropies. Entropy redistribution techniques in which a reservoir system absorbs excess entropy have been proposed \cite{bernier09, kantian18, zaletel21} and used to prepare low-entropy entangled states \cite{chiu18, yang20}, but robustly preparing many-body ground states remains challenging. 

An alternative approach is to start with an uncorrelated state, which could be prepared with very low entropy, and adiabatically transform it into a many-body quantum state. For example, quantum antiferromagnetic correlations have been observed by adiabatically loading a spin-mixture into an optical lattice \cite{boll16, parsons16, cheuk16, mitra18}. However, many such protocols require mass and entropy redistribution across the system which increases the coherence time requirements. Local transformations of the Hamiltonian have the promise of being faster and scalable to larger systems. Such protocols have been proposed \cite{rabl03, sorenson10, lubasch11, schachenmayer15, wei22} and realized \cite{pan21, sompet22} using microscopic engineering of the initial state by optical superlattices, ladder systems, or spin-dependent lattices. Finally, the importance of adiabatic preparation protocols extends beyond optical lattice systems and they have been recently utilized to prepare correlated states of quantum Hamiltonians in systems of Rydberg atom arrays \cite{ebadi21, semeghini21, ebadi22, chen22}. 

\begin{figure}[t!]
\centering
\hspace{1.5cm}
\begin{overpic}[width= 0.8\textwidth]{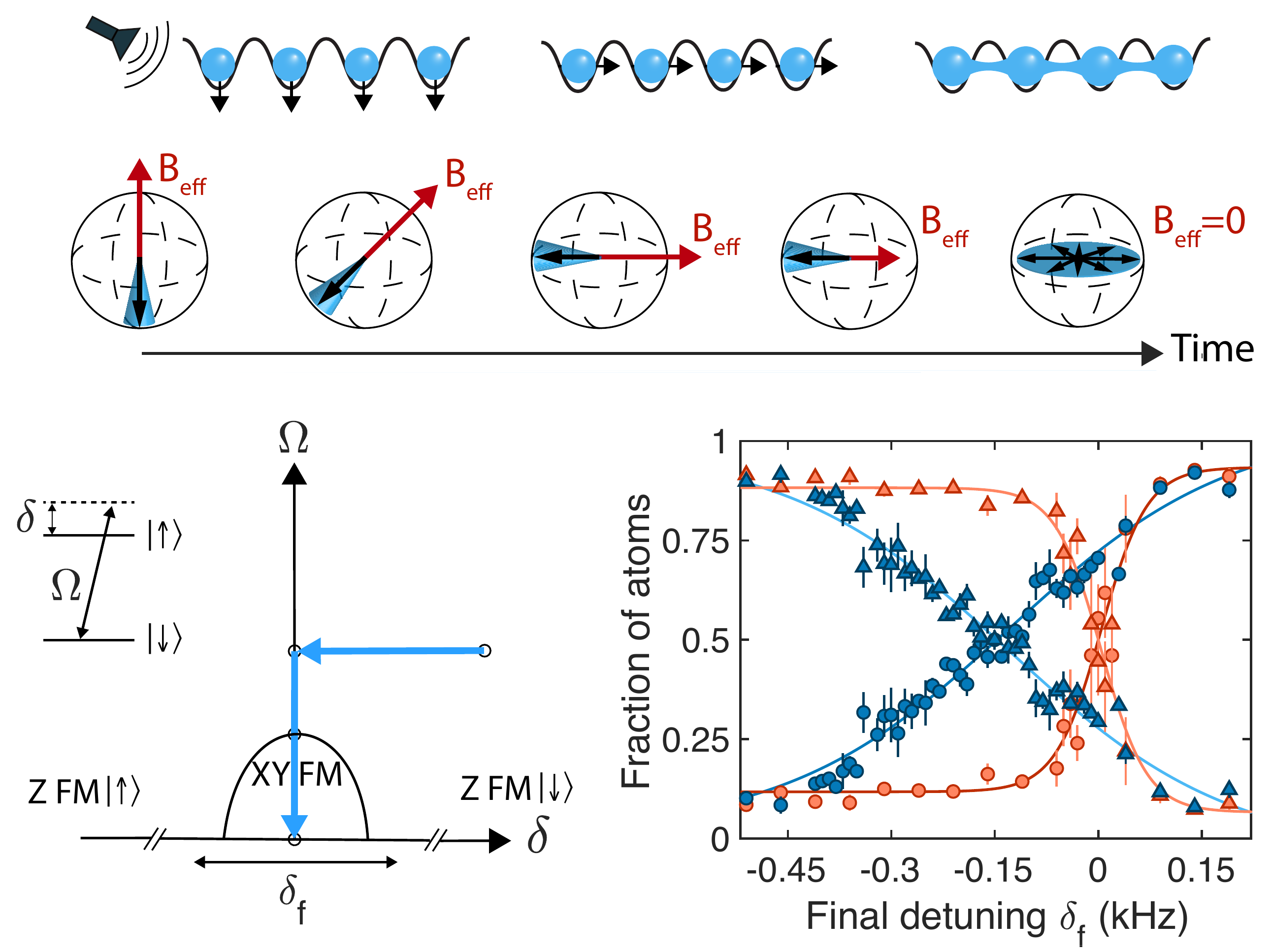}
\put(-1,72){(a)} \put(-1,39){(b)} \put(50,39){(c)}
\end{overpic}
\caption{Many-body spin rotation in 1D chains. (a) A fully magnetized state is rotated by an adiabatic passage into a correlated phase in the xy-plane which has no magnetization. 
(b) Schematic representation of the phase diagram. Starting from the z-ferromagnet (Z FM) in $|{\downarrow}\rangle^{\otimes N}$, where $N$ is the number spins, a microwave field is applied coupling the two spin states with detuning $\delta$ (effective z-magnetic field) and Rabi frequency $\Omega $ (effective x-magnetic field) with $|\delta| \gg |\Omega|$. First, the detuning is ramped to zero, rotating the spins to the xy-plane, then the Rabi frequency is ramped to zero, ideally realizing the xy-ferromagnet (XY FM). (c) Measured fraction of atoms in each state  as a function of the final detuning: $|{\downarrow}\rangle$ (circles) and $|{\uparrow}\rangle$ (triangles) for a deep 35 $E_R$ lattice of isolated sites (orange) and a shallow 11 $E_R$ lattice of coupled sites (blue). The solid lines are phenomenological fits of the form: a $tanh$(($\delta$-$\delta_0$)/w)+c. } 
\label{fig1}
\end{figure}

Here we use an adiabatic scheme which involves a direct manipulation of the spin state, and not the external potential, and requires control only over a microwave field. We demonstrate that by a many-body spin rotation, realized by an adiabatic sweep of the detuning and power of the microwave field, states with different magnetization can be connected. The properties of such rotation protocols have been explored theoretically in \cite{araceli20,araceli20b}. We realize a spin-1/2 XXZ chain in which a z-ferromagnet (a highly magnetized state) is rotated into an xy-ferromagnet, which is a strongly-correlated state with no gap in the infinite-chain limit. In a finite system, the gap is inversely proportional to the system size, allowing the adiabatic connection. The xy-ferromagnet is a magnet which points nowhere on average, i.e. it is a superposition of states which point in different directions in the xy-plane and for which the spin operator $S_z = 0$, but the expectation values are also $\langle S_x \rangle = 0$ and $\langle S_y \rangle = 0$, Fig.\ref{fig1}(a). We employ a new technique to show the presence of correlations in the many-body state: we apply a reverse microwave sweep but with a different phase relative to the initial sweep. This protocol can distinguish between isolated spins, coupled spins, and dephased spins (a collection of spins with random orientations). We recover up to 50 $\%$ of the initial magnetization independent of the phase of the reverse sweep, a strong evidence for the successful preparation of a spin state with xy-ferromagnetic correlations. The presence of correlations is further corroborated by measuring excess fluctuations in $\langle S_x^2 \rangle$, which are proportional to the Quantum Fisher information. Detailed numerical simulations verify our protocols and show that the coherence time in our system is limited by intensity noise in the microwave pulse during the final stages of the preparation when the gap is the smallest. For these timescales, our results are consistent with creating correlations over a few lattice sites. Longer chains require considerably longer time evolution to ensure adiabaticity.

\section{Experimental setup and spin Hamiltonian}

The system is a Mott insulator of $^7$Li atoms in an optical lattice. With one particle per site and two hyperfine states, it realizes the (anisotropic) spin-1/2 Heisenberg model, where effective spin-spin interactions between neighboring sites are realized by a second-order tunneling process (superexchange) \cite{altman03, ddl03}. We apply a microwave field coupling the two hyperfine states with detuning $\delta = \omega - \omega_0$, where $\hbar \omega_0$ is the energy difference between the two hyperfine states and $\omega$ is the frequency of the microwave field, and with Rabi frequency $\Omega$. This is equivalent to having a z- and an x- magnetic field in a spin system respectively, realizing the anisotropic spin-1/2 Hamiltonian with external fields:
\begin{eqnarray}
H = & J_z \sum_{\langle i,j \rangle} S_i^zS_j^z+ J_{xy} \sum_{\langle i,j \rangle}  \left(S_i^xS_j^x + S_i^yS_j^y \right) \nonumber \\
& + \delta(t) \sum_i S_i^z + \Omega(t) \sum_i S_i^x,
\label{Heisenberg}
\end{eqnarray}
where $\langle i,j \rangle$ denotes nearest neighbors, and $S^\alpha_i$ are spin operators. Here $J_{z}/h = -73.9$ Hz and $J_{xy}/h = 76.5$ Hz are the superexchange parameters, which are $\sim \tilde{t}^2/U_{\alpha\beta}$ where $\tilde{t}$ is the tunneling between neighboring sites and $U_{\alpha\beta}$ are the on-site interactions with $\alpha,\beta \in (|{\uparrow}\rangle,|{\downarrow}\rangle )$. The on-site interactions and hence the superexchange parameters can be varied by changing the applied magnetic field via Feshbach resonances (\ref{appB}).

The spins are encoded in the second-lowest and third-lowest hyperfine states $|{\downarrow}\rangle = |{m_i,m_j }\rangle = |{1/2,-1/2 }\rangle$ and $|{\uparrow}\rangle = |{-1/2,-1/2}\rangle$, respectively, at a magnetic field of 1000 G and can be imaged separately (\ref{appC}). Rather than using the lowest two hyperfine states, this encoding reduces the sensitivity to magnetic field noise by an order of magnitude. The optical lattice is formed by retroreflecting three orthogonal 1064 nm laser beams. Throughout this work we compare deep (35 $E_R$) and shallow (11 $E_R$) lattices in two configurations: i) isolated spins: all three lattices at 35 $E_R$, making the superexchange coupling between them small compared to the timescales of the experiment ($h/(4 \tilde{t}^2/U_{\uparrow\downarrow}) = 80 s$); and ii) coupled spins in 1D chains: lowering the depth of one lattice arm to 11 $E_R$ to enable tunneling, which creates a collection of spin chains with an average length of 16 sites as determined by the confining potential (\ref{appE}). 

\section{Preparation protocol}
The protocol for preparing an xy-ferromagnet using a many-body spin rotation starts with a Mott insulator of isolated spins in $|{\downarrow}\rangle$. This is the z-ferromagnetic state $|\Psi_0\rangle =|{\downarrow}\rangle^{\otimes N}$ trivially prepared by loading a Bose-Einstein condensate of $|{\downarrow} \rangle$ atoms into the lattice from an optical dipole trap. This is the highest excited state of the spin Hamiltonian \ref{Heisenberg} in the limit of large detuning $\delta \gg \Omega$. The adiabatic connection is realized at low lattice depths by performing half a Landau-Zener sweep ($\delta \rightarrow 0$) followed by an adiabatic ramp off of the driving field $\Omega \rightarrow 0$, Fig.\ref{fig1}(b). Without interactions between sites, each atom would be individually prepared in the superposition state $1/\sqrt{2} \left(|{\downarrow}\rangle + |{\uparrow}\rangle \right)$. However, nearest-neighbor interactions ($J_{xy}$) along the chain open a many-body gap in the eigenspectrum, so that the initial multi-particle state is instead adiabatically connected to an entangled state: the xy-ferromagnet. In the mean-field picture, sweeping the detuning to zero at a constant Rabi frequency rotates the spins into the xy-plane. Sweeping the Rabi frequency to zero removes the guiding x-bias field, leaving the system in the xy-ferromagnetic state which is stabilized by the spin-spin correlations, similar to the Weiss mean field. 

We first measure the effect of the adiabatic protocol on the populations in the two spin states and calibrate the resonance of the transition $|{\uparrow}\rangle \longleftrightarrow |{\downarrow}\rangle$ by varying the final point of the detuning sweep $\delta_f$. Fig.\ref{fig1} (c) shows the population going smoothly from all atoms in $|{\uparrow}\rangle$ to all atoms in $|{\downarrow}\rangle$. We denote zero detuning the point at which there is an equal number of atoms in each spin state, i.e. the total $S_z = 0$. This point is shifted for the 11 $E_R$ lattice, which is due to the non-zero tunneling at low lattice depths. From the mapping of the Bose-Hubbard Hamiltonian to the Heisenberg model, there is an additional effective z-magnetic field term $\sim \tilde{t}\,^2 (1/U_{\uparrow\uparrow} - 1/U_{\downarrow\downarrow})$ \cite{xyspiral2021}. This term is exceptionally small (and typically negligible) in a deep lattice, but in a shallow lattice it shifts the effective zero detuning point. The width of the feature in Fig.\,\ref{fig1}(c) is also larger at 11$\,E_R$ and is proportional to the coupling matrix element $J_{xy}$ between lattice sites.

\section{Probing the resulting state}

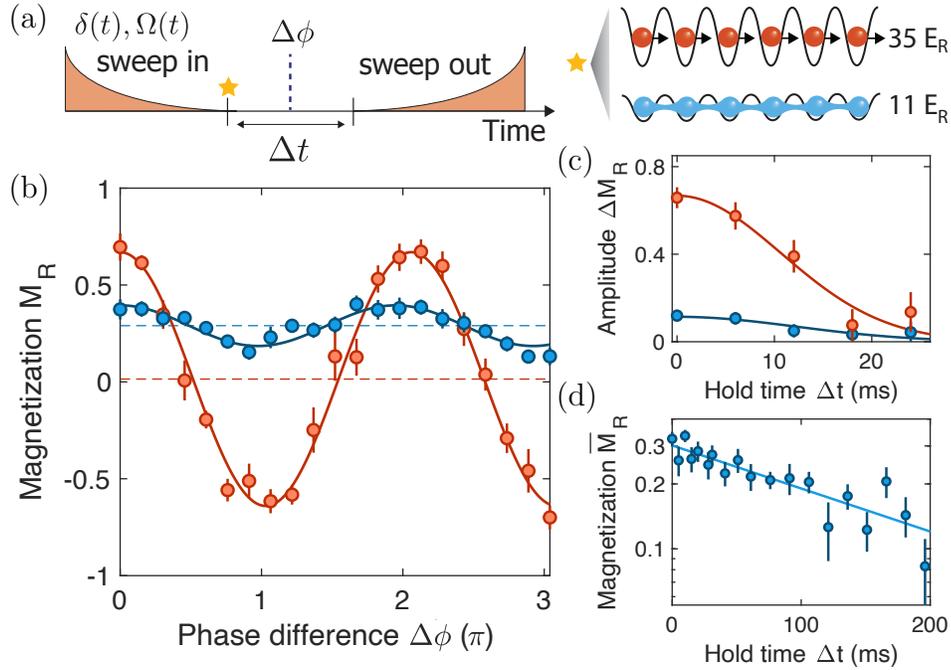
\begin{figure}[t!]
\centering
\hspace{1.5cm}
\begin{overpic}[width=0.8 \textwidth]{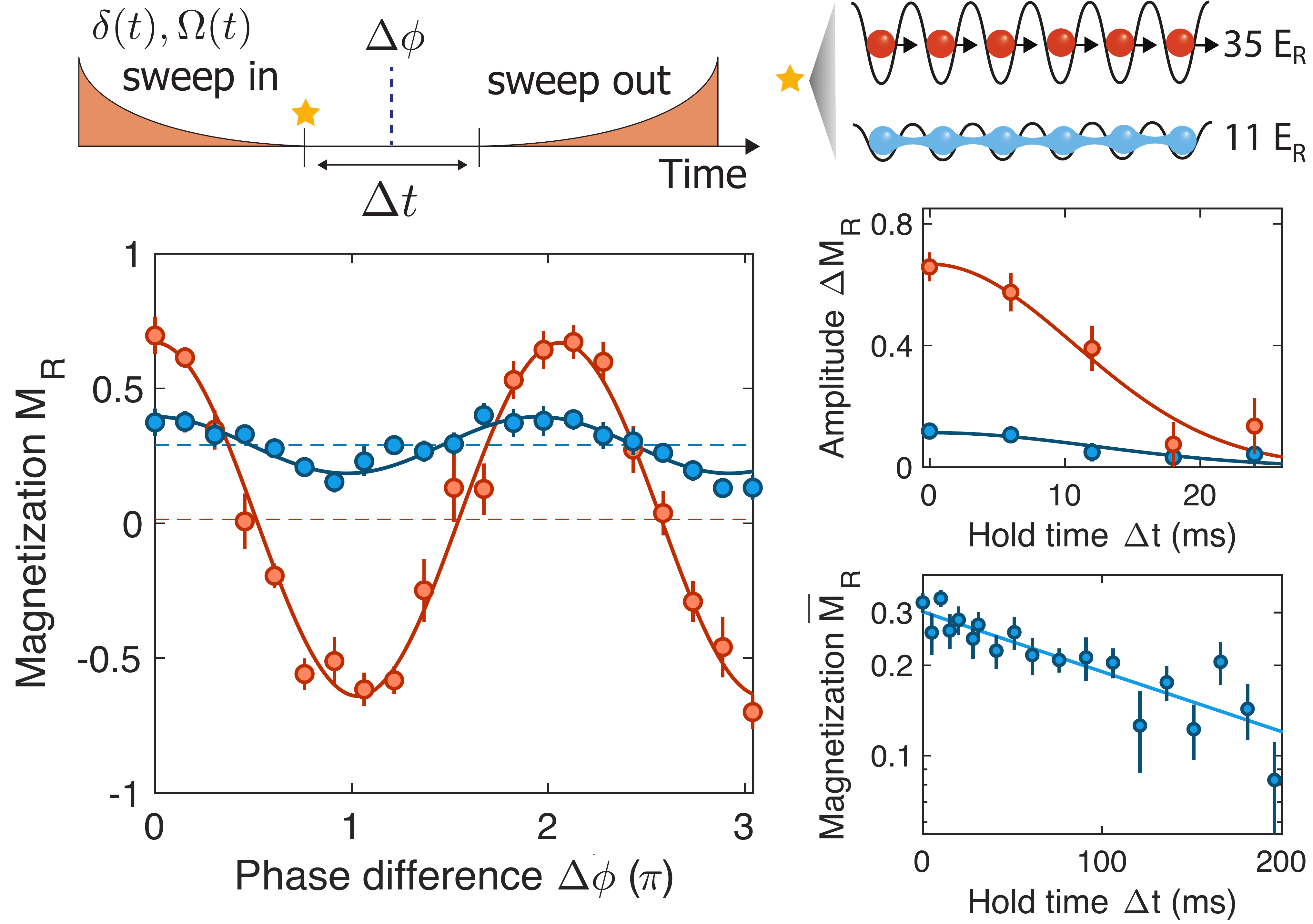}
\put(0,68){(a)}\put(0,50){(b)}\put(58,53){(c)}\put(58, 28){(d)}
\end{overpic}
\caption{Reversing the initial sweep. (a) After the initial sweep, we hold for time $\Delta t$ and apply an inverse sweep with relative phase $\Delta \phi$. (b) Return magnetization for a deep lattice (orange) and a shallow lattice (blue) for $\Delta t = 0$. The solid lines are sinusoidal fits of the form $M_R = \Delta M \cos(\Delta \phi) + \overline{M}_R$ and the dashed lines are $\overline{M}_R$. The non-zero $\overline{M}_R$ is an indication that a correlated phase related to xy-ferromagnetism has been realized in the shallow lattice. (c) Amplitude $\Delta M_R$ as a function of hold time between the sweeps. The solid lines are fits of the form $a \exp[-(t/\tau)^2]$ with $\tau_{35} = 15(5)$ ms and $\tau_{11} = 17(10)$ ms for the deep and shallow lattices respectively. (d) $\overline{M}_R$ as a function of hold time in a shallow lattice, which remains non-zero for much longer times than $\Delta M$. The solid line is an exponential fit $a\exp[-t/\tau]$ with decay time of 217(48) ms.}
\label{fig2}
\end{figure}

We perform the adiabatic sweep and use the corresponding zero-point detunings as the endpoint of the ramp for deep and shallow lattices respectively. To probe the resulting state, we implement a Ramsey-like protocol which allows us to distinguish between single-particle and correlated evolution of the spins. After performing the state preparation, we introduce a phase jump $\Delta \phi$ in the drive and then perform the sweep of the driving field in reverse, Fig.\,\ref{fig2}(a). Our observable is the return magnetization $M_R = \langle N_{\downarrow}-N_{\uparrow} \rangle/(N_{\downarrow}+N_{\uparrow})$ averaged over the cloud, which can be extracted directly from spin-sensitive images. In an ideal system of isolated spins, the state of each spin after the initial sweep has a well-defined phase and $M_R$ exhibits a Ramsey-type oscillation between $-1$ and $1$ as a function of $\Delta \phi$. In a system of coupled spins, if the protocol has successfully connected the z-ferromagnet to the xy-ferromagnet and back, we expect to measure $M_R = 1$, independent of $\Delta \phi$. Finally, if the spin rotation had instead resulted in a collection of spins with random orientations, would measure $M_R = 0$ independent of $\Delta \phi$. While a measurement of  zero-magnetization after the initial sweep could be due to the formation of a correlated phase or to dephasing, a non-zero return magnetization can emerge from a state with $S_z = 0$ after the return sweep if correlations have been established. 

The results of the measurement are shown in Fig.\,\ref{fig2}(b). We parameterize the return magnetization $M_R = \Delta M \cos(\Delta \phi) + \overline{M}_R$ by its amplitude $\Delta M$ and offset $\overline{M}_R$. For isolated spins we observe oscillations with $\overline{M}_R = 0.015(38)$ and $\Delta M \sim 0.65(5)$. We attribute the smaller than $1$ amplitude to dephasing during the sweeps, caused by technical noise, such as magnetic field noise. In the case of coupled spins (blue), we observe a non-zero $\overline{M}_R = 0.29(2)$ and a much smaller amplitude $\Delta M \sim 0.11(2)$. The residual oscillation could be due to non-adiabaticities of the sweeps and to isolated atoms at the edges of the cloud. The measured $\overline{M}_R > 0$ shows that the final state can be reversibly populated and indicates the formation of correlations within the spins in each chain. 
 
The dependence of $M_R$ on the hold time $\Delta t$ between the initial and reverse sweep reveals the different sensitivity of the isolated and coupled spins to noise sources. The amplitude $\Delta M_R$ decays on similar timescales in both a deep and a shallow lattice, shown Fig.\,\ref{fig2}(c). The oscillations have dephased after $\sim$ 15 ms, consistent with magnetic field noise on the $10^{-5}$ level affecting isolated spins. By contrast, we expect the correlations in the coupled system to be insesitive to this level of magnetic field noise and we observe that $\overline{M}_R$ remains non-zero for longer, Fig.\,\ref{fig2}(d). Finally, for hold times longer than 150ms we measure $10 \%$ atom loss, possibly due to lattice heating or spin-changing collisions.

\section{Improving the return magnetization $\overline{M}_R$}

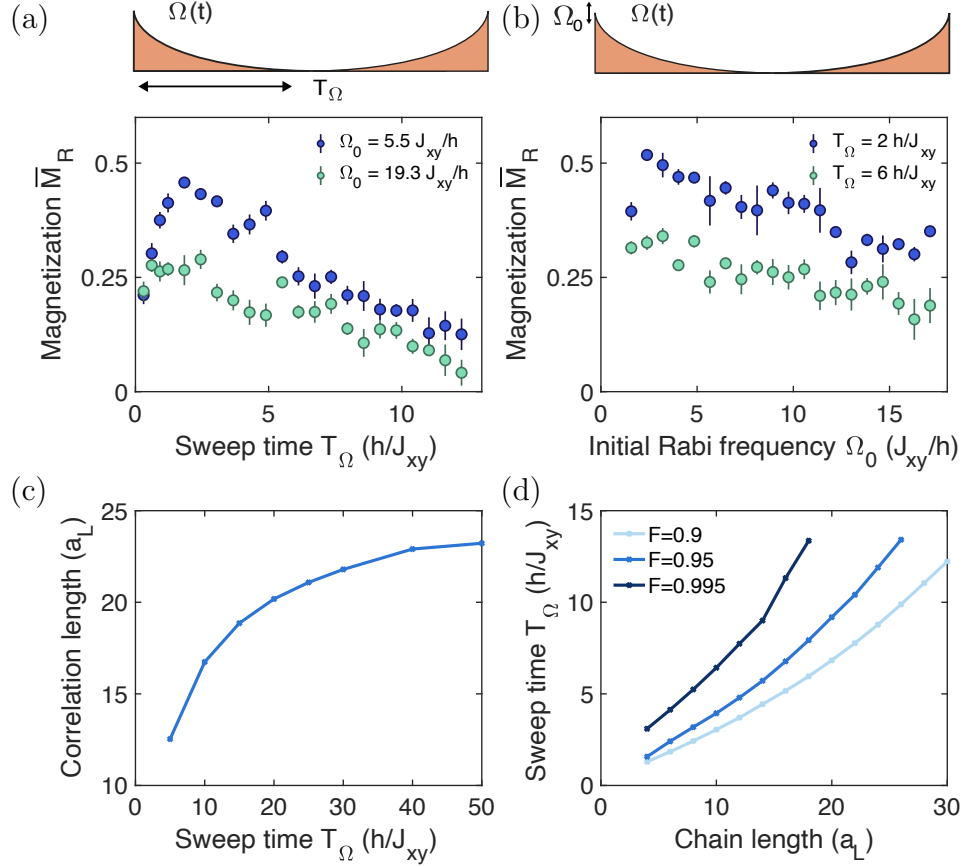
\begin{figure}[t!]
\centering
\hspace{1.5cm}
\begin{overpic}[width=0.8\textwidth]{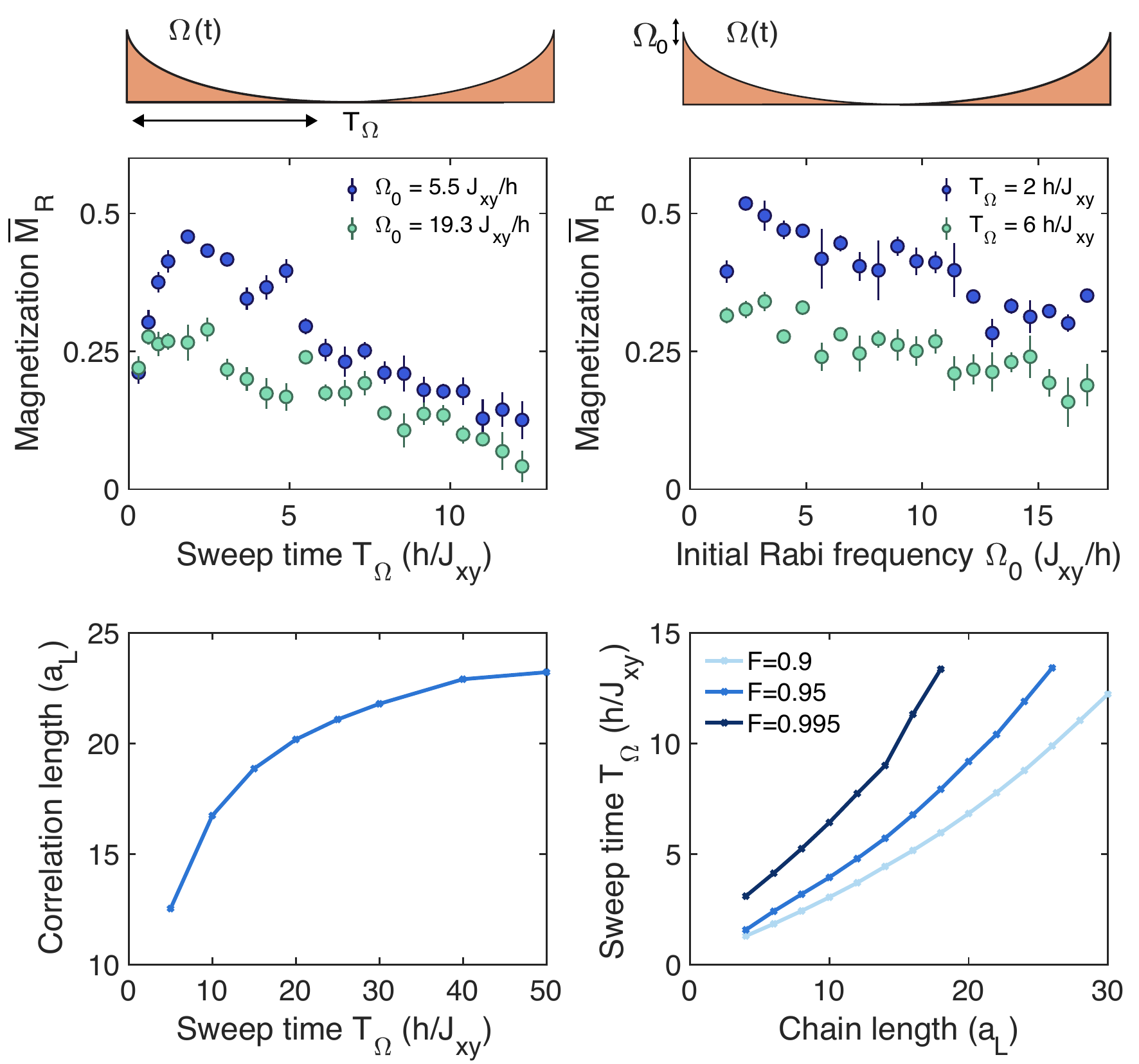}
\put(-2,89){(a)}
\put(50,89){(b)}
\put(-2,39){(c)}
\put(50,39){(d)}
\end{overpic}
\caption{Optimization of sweep parameters. The return magnetization $\overline{M}_R$ can be improved by varying: (a) the sweep time $T_{\Omega}$ of the Rabi frequency and (b) the initial Rabi frequency $\Omega_0$. (c) Numerical simulations of the ideal preparation scheme for $N = 100$ sites. The correlation length $\eta$ is extracted by an exponential fit $A\exp(-\eta m)$ to the off-diagonal spin correlation function $\langle \hat{S}^+_{N/2} \hat{S}^-_{N/2+m}\rangle$. (d) Ramp time $T_{\Omega}$ required for the fidelity $F=|\langle \psi_{\rm prep}(T_{\Omega})|\psi_{GS}\rangle|^2$ to reach a certain threshold as a function of chain length. Here $|\psi_{GS}\rangle$ is the ground state, $|\psi_{\rm prep}\rangle$ is the prepared state and $a_L$ is the lattice spacing. }
\label{fig3}
\end{figure}

The non-oscillating return magnetization $\overline{M}_R$ is a measure of the fidelity of the preparation of the target state and can be used to optimize the sweep parameters. We observe that $\overline{M}_R$ can be increased by using lower initial Rabi frequency $\Omega_0$ and shorter Rabi frequency sweeps. This is plotted in Fig.\,\ref{fig3}. $\overline{M}_R$ reaches a maximum of 0.51 for $\Omega_0 \sim 5 J_{xy}/h = 382$ Hz and for a one-way Rabi frequency sweep time of $T_\Omega \sim 2 h/J_{xy} = 26$ ms. In principle, the longer the sweep timescale, the better the adiabaticicy and therefore the fidelity of the preparation. Numerical simulations show that in a system of 100 sites, the correlation length increases logarithmically with sweep time as seen in Fig. \ref{fig3}(c), making the preparation of fully correlated long chains challenging. The required time to reach a certain fidelity as a function of chain size is plotted in Fig. \ref{fig3}(d). For chain lengths of 15-20 sites, as used here, the ramp times for the Rabi frequency sweep required to reach a fidelity of 0.9 are 6-7 $h/J_{xy}$. This corresponds to correlation lengths of about 13 sites and return magnetization of more than 0.90. The experimental values are lower.  This and the fact that there is a maximum in the observed $\overline{M}_R$ as a function of ramp time points to the presence of technical noise in the experiment leading to dephasing. Numerical simulations of various sources and levels of technical noise suggest that the main source of noise affecting the fidelity of the preparation is intensity noise of the microwave pulse during the final stages of the sweep (\ref{appG}). For example, for a Rabi frequency of 0.2 $J_{xy}/h$ the coherence time of single-particle Rabi oscillations in a deep lattice is $\sim 1.5\, h/J_{xy}$, allowing for a single superexchange event.

\section{Quantum Fisher Information}
\begin{figure}[t!]
\centering
\hspace{1.5cm}
\begin{overpic}[width=0.8\textwidth]{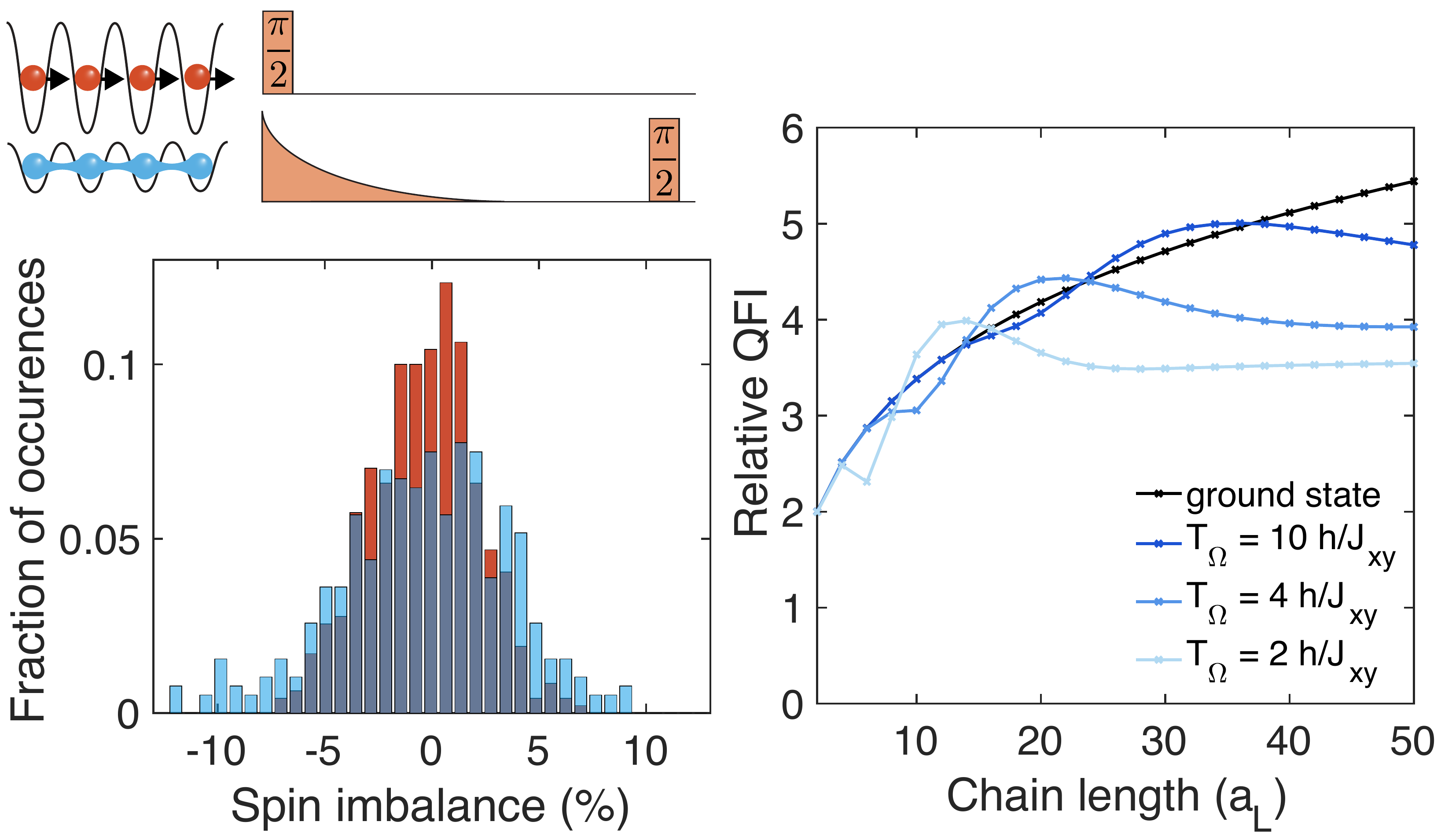}
\put(-5,53){(a)}
\put(50,53){(b)}
\end{overpic}
\caption{Quantum Fisher Information. (a) Histograms of the spin imbalance $I$ for isolated sites rotated to the xy-plane (orange) and for coupled sites (blue). The protocol in each case is shown above. The variance of $S_x$, as measured by the variance of $I$, is 2.66 times larger for coupled sited as compared to the shot-noise-limited variance of single sites, indicating the presence of correlations at low lattice depths. Here $T_\Omega = 3.2 \, h/J_{xy}, \Omega_0=5.5 \, J_{xy}/h$. (b) Numerical simulations showing the QFI for pure states, ${\rm QFI} = 4 \langle \hat{S}_x^2 \rangle - 4\langle \hat{S}_x \rangle^2$ as a function of chain length for different ramp times. Values are given relative to the QFI for independent spins.}
%
\label{fig4}
\end{figure}

A way to probe the correlated phase without the reverse ramp is to measure the variance of the spin operator $S_x = \sum_i S^x_i$, where the sum is over lattice sites $i$. In the case that we assume the state to be pure, we note that the variance is proportional to the Quantum Fisher Information (QFI) in this system, which can be used to quantify many-body entanglement \cite{araceli20b}. When single spins are rotated to the xy-plane, the variance of $S_x$ is shot noise-limited: $\langle S_x^2\rangle \propto N$, where $N$ is the total number of spins. By contrast, the presence of correlations in a coupled system render it delocalized in the xy-plane (i.e. spins do not ``point'' in a particular direction in the xy-plane), so that a measurement of $S_x$ should exhibit larger fluctuations, compared to shot noise. The variance of $S_x$ can be measured by applying a $\pi/2$ pulse after the initial sweep, which maps $S_x$ to $S_z = N_{\uparrow} - N_{\downarrow}$. The statistics of the spin imbalance $I = \langle N_{\uparrow}-N_{\downarrow} \rangle/(N_{\uparrow}+N_{\downarrow})$ are shown in Fig.\,\ref{fig4}(a) for the coupled system, compared to a system of isolated spins rotated to the xy-plane. While the standard deviation of the latter is measured to be given by shot noise, we find that the variance of the spin imbalance is larger for coupled spins by a factor of $\langle S_x^2\rangle_{\rm coupled}/\langle S_x^2\rangle_{\rm isolated} = 2.66$. The predicted QFI relative to the QFI of single spins as a function of chain length is plotted in Fig.\,\ref{fig4}(b). The increased variance of $S_x$ measured here corresponds to a relative QFI of 2.66 and corroborates the existence of correlations over a few sites.

\section{Conclusions}
Our combined experimental and theoretical study demonstrates the potential of adiabatic spin rotation for creating new many-body quantum states. The comparison of experimental and numerical results provided guidance for optimized sweep parameters, and allowed us to identify which sources of noise limited the fidelity of the state preparation.  The calculations also show that the fidelity depends drastically on the chain length.  In our current system, we average over an ensemble of chain lengths.  A major improvement would be the use of a quantum gas microscope where chains of specific lengths can either be prepared or post-selected.  In addition, the effect of holes in chains could be characterized. Longer correlated states could be created by extending the coherence timescale by improving the stability of the microwave field and the magnetic field and by using defect-free initial Mott insulating states. 

Our results showcase adiabatic passage protocols for preparing correlated quantum phases. With improved detection methods, our system can be used to study the properties of entangled many-body states. As an example, in the limit $J_z/J_{xy} \rightarrow -1$ the QFI of the xy-ferromagnet is maximized with possible applications in quantum sensing.   Our protocol can be extended to preparing other many-body states since the anisotropy of the spin Hamiltonian can be widely varied. For example, the xy-antiferromagnet can also be prepared through adiabatic spin rotation by including a magnetic field gradient which is ramped adiabatically. In addition, our platform can be  used to develop other state preparation protocols, e.g. counter-diabatic driving \cite{berry09,demirplak03,ieva22},  which are faster than adiabatic ramps and possibly superior when technical noise limits the preparation time.

\ack{We thank Araceli Venegas-Gomez and Johannes Schachenmayer for useful discussions. We acknowledge support from the NSF through the Center for Ultracold Atoms and Grant No. 1506369, the Vannevar-Bush Faculty Fellowship, and DARPA. Y. K. L. is supported in part by the National Science Foundation Graduate Research Fellowship under Grant No. 1745302. Work at the University of Strathclyde was supported by the EPSRC Programme Grant DesOEQ (Grant No. EP/P009565/1), and by AFOSR Grant No. FA9550-18-1-0064.}

%
\section*{References}
\bibliographystyle{iopart-num}
\bibliography{asp_bibliography}

\appendix{}
\section{Hamiltonian}\label{appA}
Spin Hamiltonians can be realized with ultracold bosons in optical lattices in the Mott insulator state using the tunneling between lattice sites $\tilde{t}$ and on-site interactions $U$ \cite{altman03}. Here we use a Mott insulator with one atom per site and two hyperfine states, which encodes the anisotropic spin-1/2 Heisenberg model, which we have implemented before \cite{ivana20, nature20}:
\begin{eqnarray}
H = & J_z \sum_{\langle i,j \rangle} S_i^zS_j^z+ J_{xy} \sum_{\langle i,j \rangle}  \left(S_i^xS_j^x + S_i^yS_j^y \right) \nonumber
\label{Heisenberg_supp}
\end{eqnarray}
where the sums are over nearest-neighbors. The spin parameters are:
\begin{eqnarray}
J_z& = \frac{4\tilde{t}^2}{U_{\uparrow\downarrow}} - \frac{4\tilde{t}^2}{U_{\uparrow\uparrow}} -\frac{4\tilde{t}^2}{U_{\downarrow\downarrow}} \nonumber \\
J_{xy}&=-\frac{4\tilde{t}^2}{U_{\uparrow\downarrow}} 
\end{eqnarray}
and the spin matrices $S^{\alpha}_i$ are defined as $S^z_i\,{=}\,(n_{i\uparrow}\,{-}\,n_{i\downarrow)}/2$, $S^x_i\,{=}\,(a^{\dagger}_{i\uparrow}a_{i\downarrow}\,{+}\,a^{\dagger}_{i\downarrow}a_{i\uparrow})/2$, and $S^y_i\,{=}\,{-}\,i( a^{\dagger}_{i\uparrow}a_{i\downarrow}\,{-}\,a^{\dagger}_{i\downarrow}a_{i\uparrow})/2$.

In this model, the xy-ferromagnet is the highest excited state in the range $-1<J_z/J_{xy}<1$. The gap to the nearest state increases smoothly when the anisotropy is varied from $J_z/J_{xy} \rightarrow 1$ to $J_z/J_{xy} \rightarrow - 1$. Also, the state itself varies in that range but it remains in the realm if xy-ferromagnetism. For technical reasons, we took the data for Fig.2 in the main text at 1025 G, where $J_z/J_{xy}=-0.15$ (with $J_z/h = -12.8$ Hz and $J_{xy}/h = 88.7$ Hz) and the data for all other figures at  1000 G where $J_z/J_{xy}=-0.97$ with ($J_z/h = -73.9$ Hz and $J_{xy}/h = 76.5$ Hz). Since the gap is bigger at the latter point, we expect our state preparation to work better there. However, no significant difference in the return magnetization $\overline{M}_R$ was observed.

\begin{figure}[h!]
\centering
\begin{overpic}[width=0.65\textwidth]{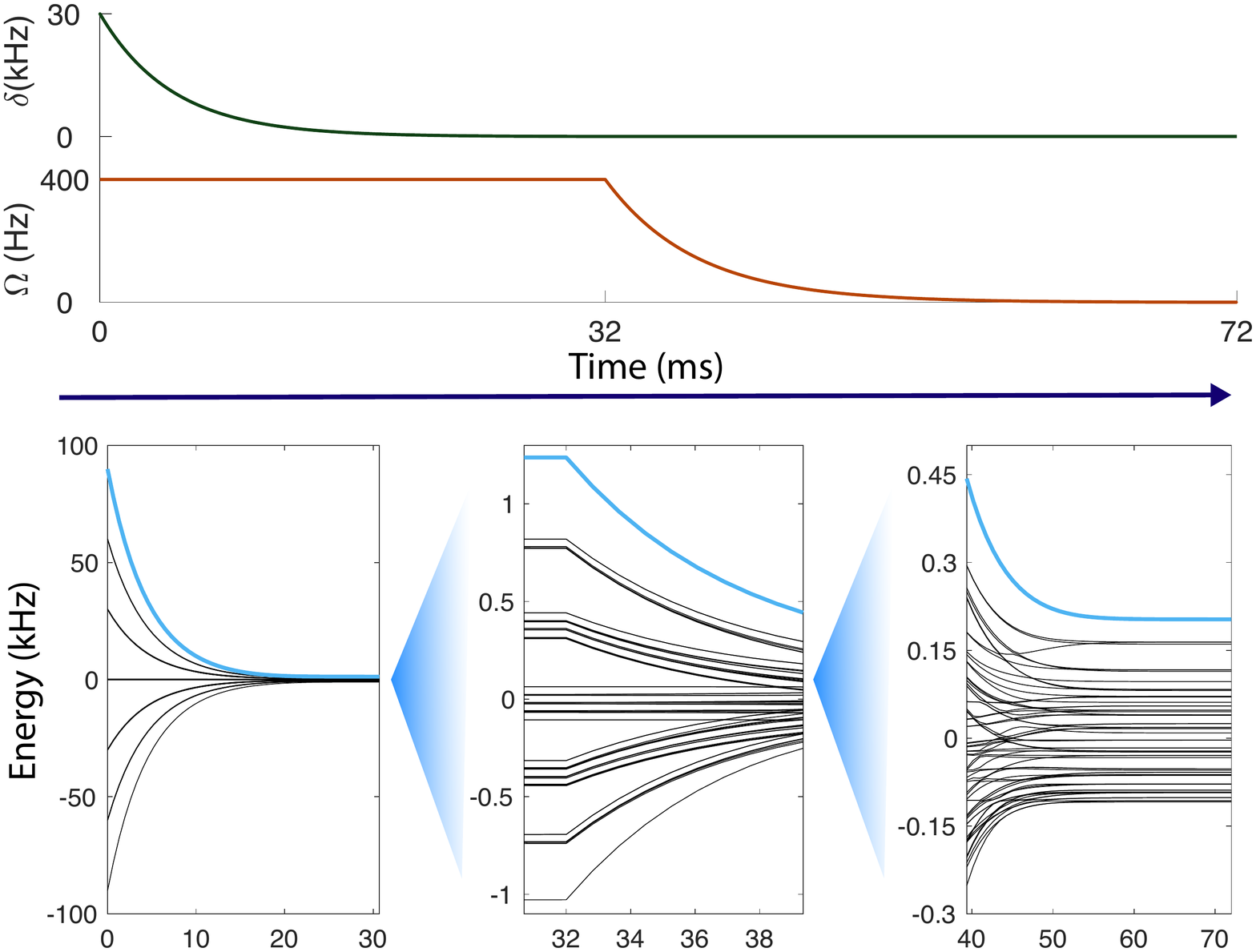}
\put(-6,74){(a)}
\put(-6,40){(b)}
\end{overpic}
\caption{Adiabatic sweep. (a) The sweep of the detuning $\delta(t)$ and Rabi frequency $\Omega(t)$ of the microwave drive between $|{\uparrow}\rangle$ and $|{\downarrow}\rangle$. (b) The evolution of the energy level diagram, shown here for a spin chain of 6 sites, highlighting the highest excited state which we follow during the sweep.}
\label{s1}
\end{figure}

The evolution of the energy level diagram as a function of sweep time is illustrated in Fig.\ref{s1} for $J_z/J_{xy}=-0.88$. Note that the gap decreases as a function of time for this protocol and the smallest gap is at the end of the sweep.

\section{Choice of spin states}\label{appB}
The spin parameters $J_z$ and $J_{xy}$ can be varied by the $^7$Li Feshbach resonances in the region 500-1500 G. In this region, the lowest 4 hyperfine states with $m_J = -1/2$ could be suitable choices of spin states. Typically, the lowest two have been used to realize spin models. However, here we use the second and third lowest states with $|{m_I}\rangle = -1/2$ and $|{m_I}\rangle = 1/2$ due to their lower sensitivity to magnetic field noise. These states have a very small relative magnetic moment $|\mu_{\downarrow}-\mu_{\uparrow}|$ = 2.76 kHz/G, compared to $\sim$ 30kHz/G for the lowest two hyperfine states. The magnetic field noise in our system is $\sim$ 3.5 mG, corresponding to stability at the $10^{-5}$ level, and resulting in 10 Hz noise, which is $\sim$ 7.5 times smaller than the superexchange timescale. 

\begin{figure}[h!]
\centering
\begin{overpic}[width=0.9\textwidth]{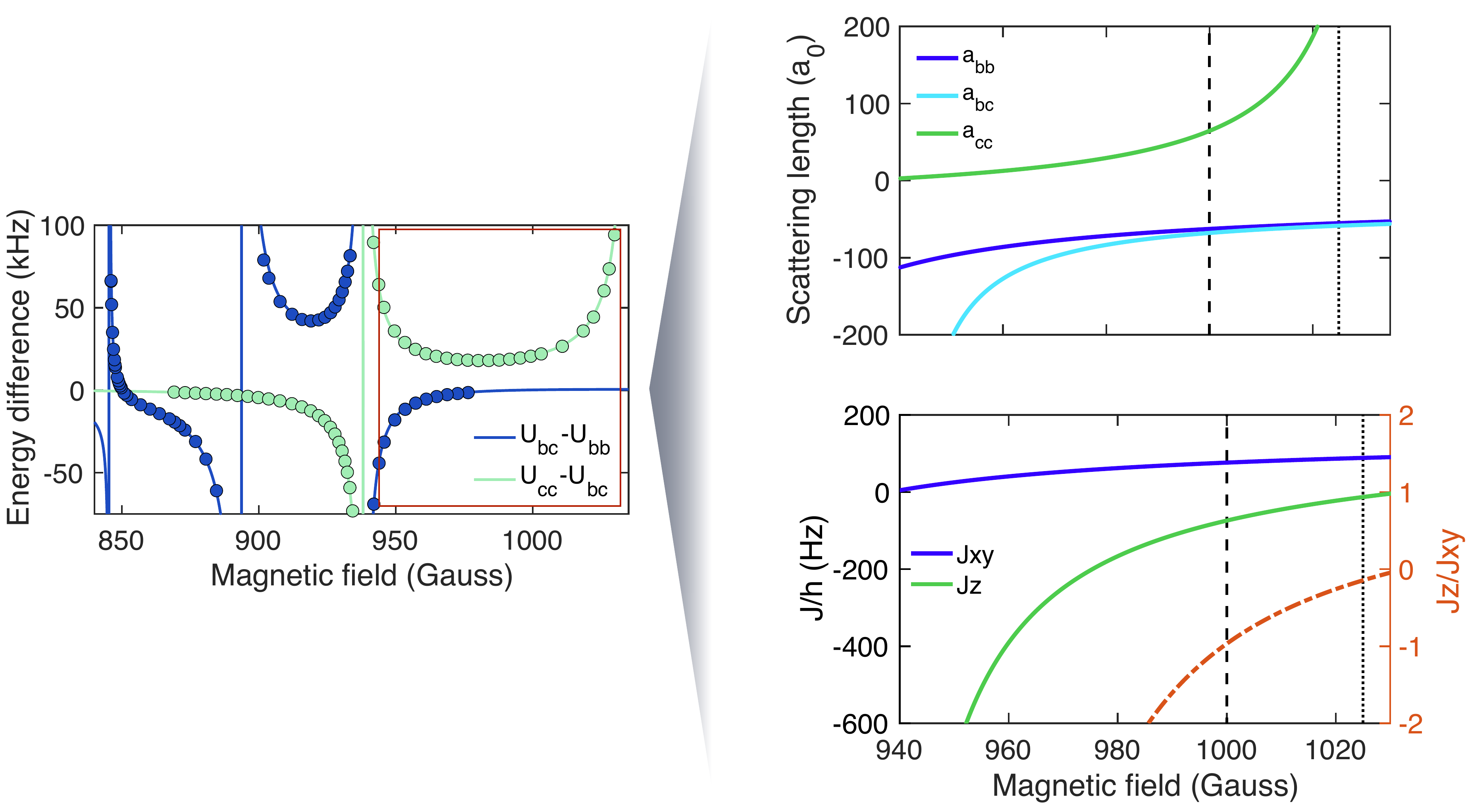}
\put(-5,39){(a)}
\put(49,53){(b)}
\put(49,26){(c)}
\end{overpic}
\caption{Feshbach resonances in $^7$Li for the $b$ and $c$ states. (a) Energy differences $U_{bc} - U_{bb}$ and $U_{cc} - U_{bc}$ as measured by interaction spectroscopy of an n=2 Mott Insulator at a lattice depth of 35 $E_R$. (b) Scattering lengths as a function of magnetic field. (c) Parameters of the XXZ Hamiltonian as a function of magnetic field. The dashed line at 1000 G indicates the point where the data is taken except for the data in Fig.2 of the main text, which is taken at 1025 G (dotted line).   }
\label{s2}
\end{figure}

To determine the scattering lengths for these energy levels, we use interaction spectroscopy, as in \cite{jesse19}, to measure the energy differences $U_{bc}-U_{bb}$ and $U_{cc}-U_{bc}$, where we use spectroscopic notation, shown in Fig.\ref{s2}(a). The scattering length as a function of magnetic field $B$ can be approximated as a parabola:
\begin{equation}
    a(B) = a_{\rm bg} \left(1 - \sum_i\frac{\Delta_i}{B - B_{i,0} } \right)
\end{equation}
where $a_{\rm bg}$ is the background scattering length, $B_{i,0}$ are the magnetic fields of the Feshbach resonances and $\Delta_i$ are the widths of the resonances. Using the data for the ${bb}$ channel from \cite{jesse19}, we can determine the parameters for the $bc$ and $cc$ channels. This is summarized in Table \ref{tab:fbres}. 
\begin{table}[h!]
    \centering
    \begin{tabular}{ || c  c  c c ||}
    \hline
          Channel & $a_{\rm bg}/a_0$ & $\Delta$ (G) & $B_0$ (G) \\
            \hline
          bb \cite{jesse19} & $-23.0(1.4)$ & $- 14.9(0.9)$ & 845.45(02)   \\
          bb \cite{jesse19} & $-23.0(1.4)$ & $- 172.7(10.0)$ &  893.84(18) \\
          bc &  $- 35.4(2.3)$ & $- 56.9(3.7)$  &  938.11(0.05) \\
          cc  & $- 34.3(4.9)$ & $- 104.3(10.4)$ &  1036.19(0.56) \\
          \hline
         
    \end{tabular}
    \caption{Feshbach resonance parameters for the $b$ and $c$ states of $^7$Li from interaction spectroscopy in a 3D Mott insulator at 35$\,E_R$.}
    \label{tab:fbres}
\end{table}

The scattering lengths of the relevant hyperfine states are plotted in Fig.\,\ref{s2}(b) and the corresponding spin parameters are plotted in Fig.\,\ref{s2}(c).

\section{State-selective imaging}\label{appC}
\begin{figure}[h!]
\centering
\begin{overpic}[width=0.45\textwidth]{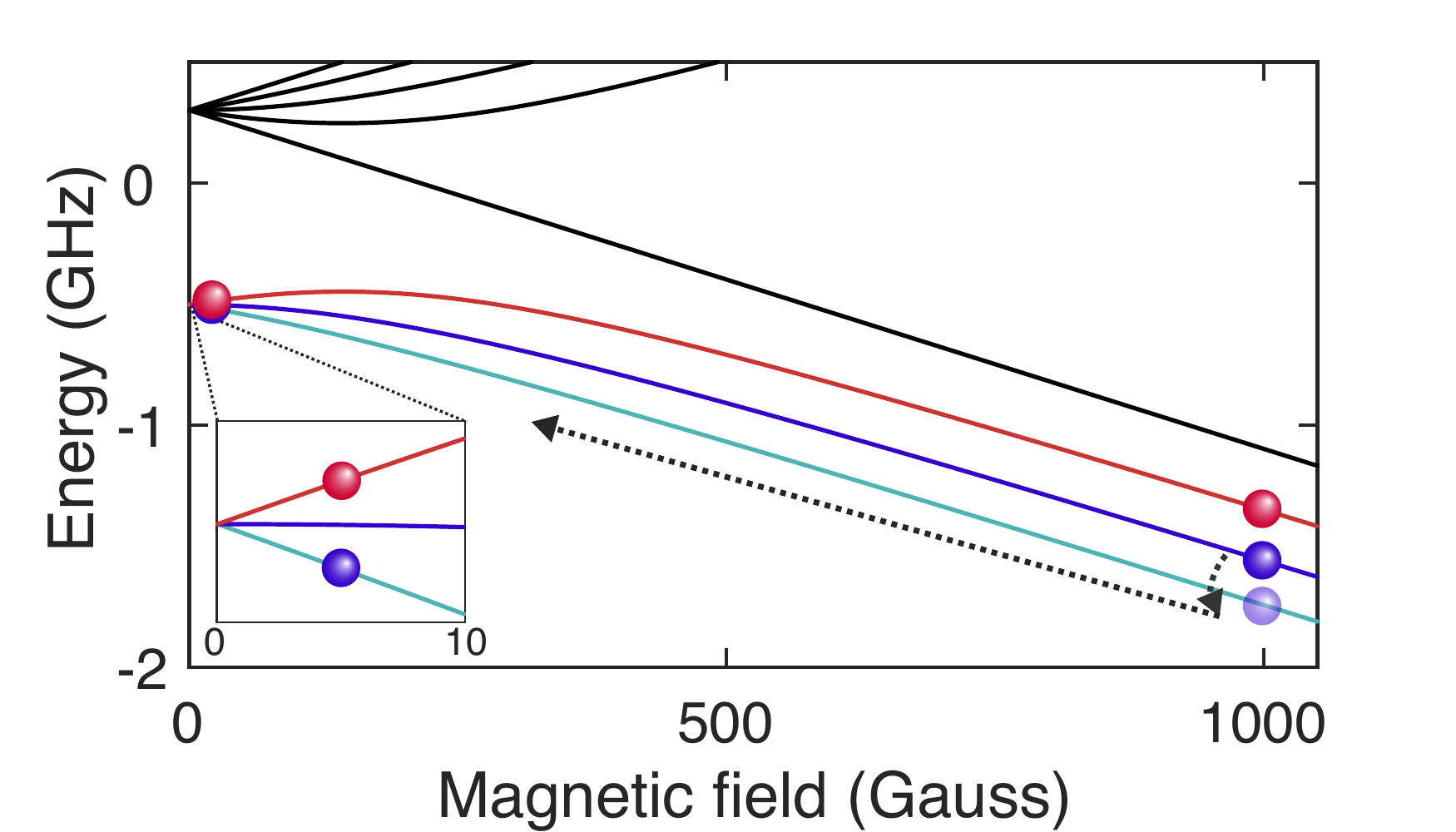}
\end{overpic}
\caption{Stern-Gerlach imaging. First, at high field, the population in the $|{\downarrow}\rangle$ (blue) is transferred to the lowest hyperfine state by a Landau-Zener transfer. Then, the field is lowered to $\sim$ 5 G, where the differential magnetic moment between the two states is large. A magnetic field gradient separates the atoms in the two states after the lattice depths are ramped down. }
\label{s2b}
\end{figure}

In this paper we use two different imaging techniques.  For the data in Fig.2, we use standard absorption imaging, in which the two states are imaged separately, since the imaging frequencies differ by $\sim$ 200 MHz. This requires repeating the experimental sequence in order to image each state, which requires longer experimental times and is sensitive to shot-to-shot atom number fluctuations. Therefore, for the data in all other figures, we implemented a more efficient technique, using Stern-Gerlach imaging, in which the two states are separated in space and can be imaged at the same time. Since the spin states have similar magnetic moments at high field, in order to separate them spatially, we map them to their low-field counterparts. We transfer the population in the $|{\downarrow}\rangle = {|1/2,-1/2}\rangle$ to $|{a}\rangle = |{3/2,-1/2}\rangle$ via a Landau Zener sweep (Fig. \ref{s2b}). This is possible because the energy differences between the different pairs of hyperfine states at these magnetic fields are significantly different, so that the different transitions can be spectroscopically distinguished. Now the two states map to the low-field states $|{a}\rangle \rightarrow |{F, m_F}\rangle = |{1,-1}\rangle$  and $|{\uparrow} \rangle \rightarrow |{F, m_F}\rangle = |{1,1}\rangle$, which have a relative magnetic field moment of 1.4 MHz/G. 

In order to measure the populations in each of these states, we quickly ramp all lattice arms to 35$E_R$, lower the magnetic field in 10 ms to about 5 G. We apply a magnetic field gradient of 50 G/cm, lower the lattice arm in the direction of the magnetic field gradient to 0 and the other two arms to 13$E_R$ and let the atoms expand. This results in two spatially separated clouds corresponding to the original spin states. We calibrate the relative number of atoms in the spin states by driving Rabi oscillations between the two spin states at high and at low fields.  The oscillation amplitude for the two coupled spin states corresponds to the same atom number.

\section{Sweep parameters}\label{appD}
\begin{figure}[h!]
\centering
\begin{overpic}[width=0.45\textwidth]{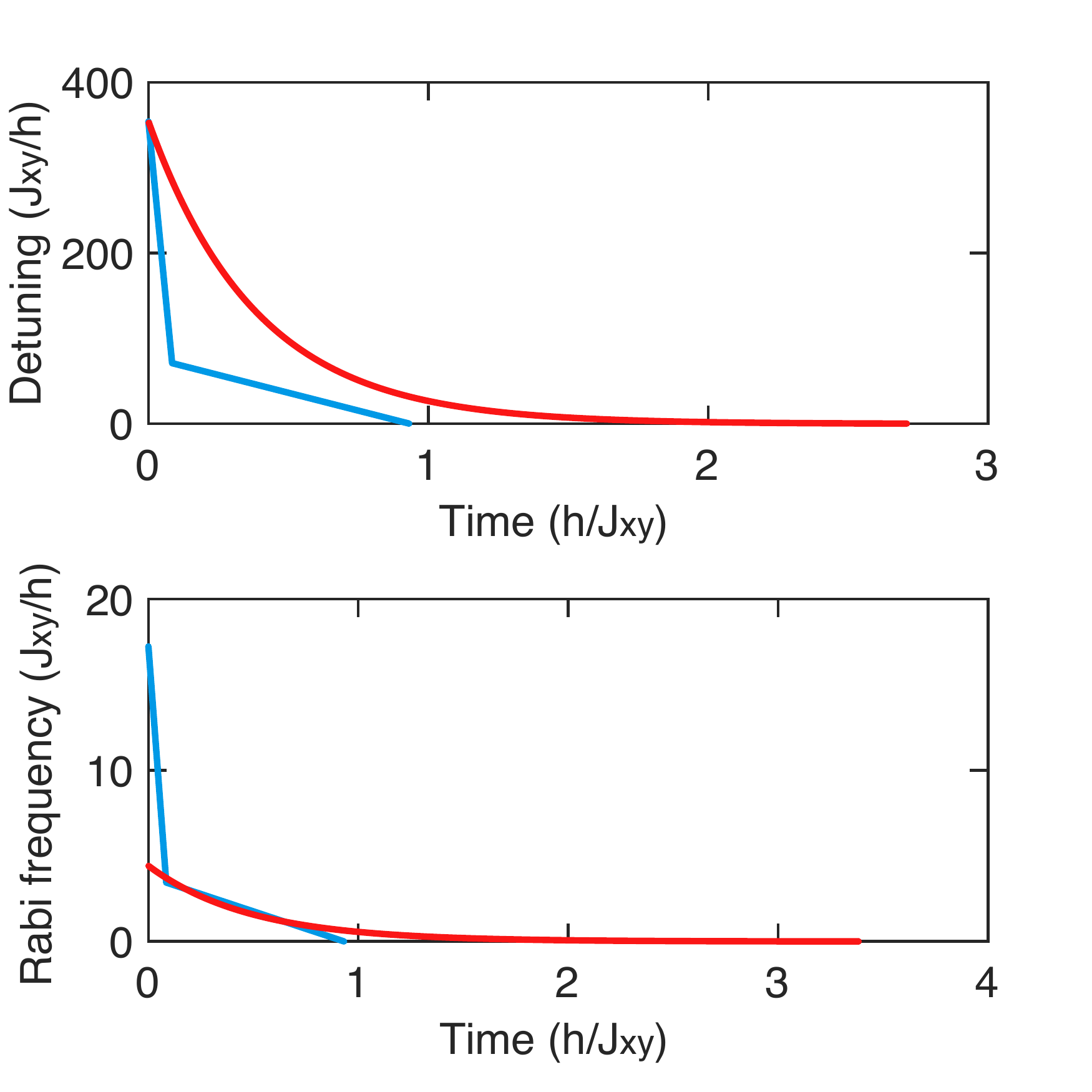}
\put(-8,92){(a)}
\put(-8,43){(b)}
\end{overpic}
\begin{overpic}[width=0.45\textwidth]{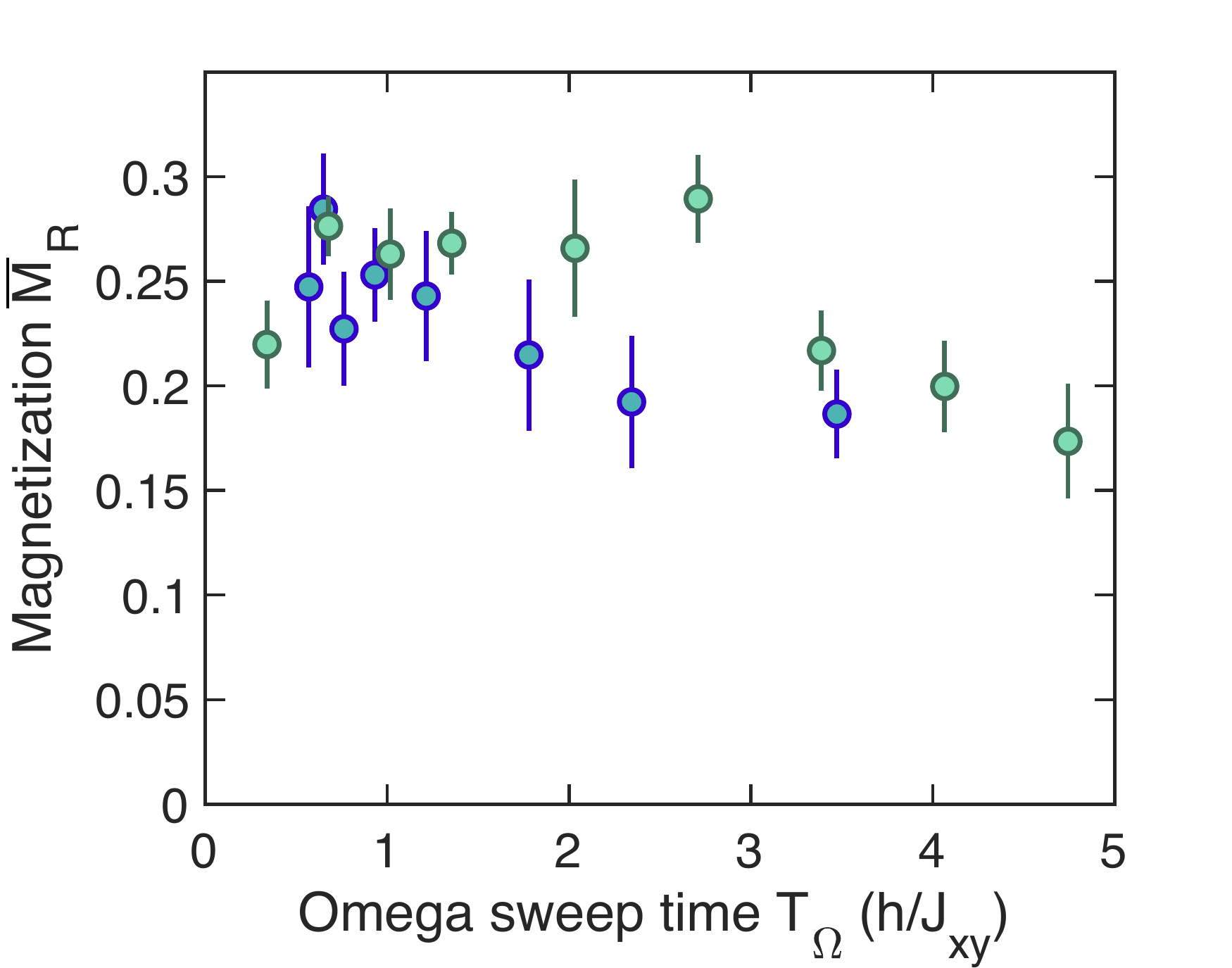}
\put(-4, 70){(c)}
\end{overpic}
\caption{Sweep parameters: (a) detuning and (b) Rabi frequency sweeps showing the optimized piece-wise linear sweep used in Fig.2 in the main text (blue) and the optimized exponential sweep used in all other figures (red). (c) Dependence of the return magnetization $\overline{M}_R$ on the total length of the Rabi frequency sweep for (blue) piece-wise linear sweeps and (green) exponential sweeps (green data in Fig.3(a) in the main text). }
\label{s3}
\end{figure}

We explored two types of sweeps: piece-wise linear (used for the data in Fig.2) and exponential (used for all other figures). We find no significant difference between the two sweeps when the timescales of the two are matched. The optimized linear sweeps and the optimized exponential sweeps are plotted in Fig. \ref{s3}(a-b). 

Fig.\ref{s3}(c) shows the average return magnetization $\overline{M}_R$ as a function of the length of the Rabi frequency sweep for both piece-wise linear and exponential pulses. In both cases we start with the maximum Rabi frequency $\Omega_0 =  19.3 \, J_{xy}/h$. For the piece-wise linear sweeps only the length of the second linear part is varied. The return magnetization is about 0.3 in both cases. 

\section{Distribution of chain lengths}\label{appE}
In order to achieve deep lattices, we focus the lattice beams to 125 $\mu m$ $1/e^2$ radius. This curvature leads to a considerable trapping potential which gives the Mott insulator a spherical shape. This leads to a distributions of chains with different lengths and to some isolated atoms at the edges of the sample. As we have discussed in Supplementary Fig. 10 of \cite{nature20}, this results in the following distribution: For $N = 6000$ atoms in the Mott insulator, the maximum chain length is $L_{max} = 21 a_L$, where $a_L=532$ nm is the lattice spacing. The average chain length is $L_{avg}=(3/4) L_{max}=16a_L$ and the total number of chains is $\pi(L_{max}/2a_L)^2 = 350$. This distribution of chain lengths complicates the optimization of the adiabatic protocol since its performance depends on chain length.

\section{Characterization of isolated particles}\label{appF}
A source of error in the measurement of the return magnetization is the presence of isolated particles at low lattice depths. These could be thermal atoms from imperfect state preparation or isolated particles at the edges of the cloud, where, due to the lattice curvature, the energy difference between neighboring sites $\Delta > 4 \tilde{t}$. Since they behave as single particles, their presence will artificially increase the return magnetization at low lattice depths. This is due to the different detunings at which $\langle S_z \rangle = 0$, i.e. the spins are in the xy-plane, which is due to the fictitious magnetic field at low lattice depths (Fig 1(c)). At deep lattices for single spins $\langle S_z \rangle = 0$ for $\delta^{(s)}_f = 0$ and at 11$\,E_R$ for coupled spins $\delta^{(c)}_f = -0.15$ kHz. Therefore, for ideal sweeps, $\overline{M}_R \rightarrow 1$ for single spins at the resonance of the coupled spins $\delta^{(c)}_f$. 

This effect can be used to estimate the fraction of single particles at low lattice depths by measuring the return magnetization $M_{R}(\delta = 0)$ at low lattice depths at $\delta^{(s)}_f = 0$. $M_{R}(0)$ is the sum of $M^{(s)}_{R}(0) = 0$ for single particles and $M^{(c)}_{R}(0) \neq 0$ for coupled spins. More generally:
\begin{equation}
 M_{R}(\delta) = \alpha_s M^{(s)}_{R}(\delta) + (1-\alpha_s) M^{(c)}_{R}(\delta)  
 \label{eq:mr_delta}
\end{equation}
where $\alpha_s$ is the fraction of single particles present. 

\begin{figure}[h!]
    \centering
    \includegraphics[width= 0.5\linewidth]{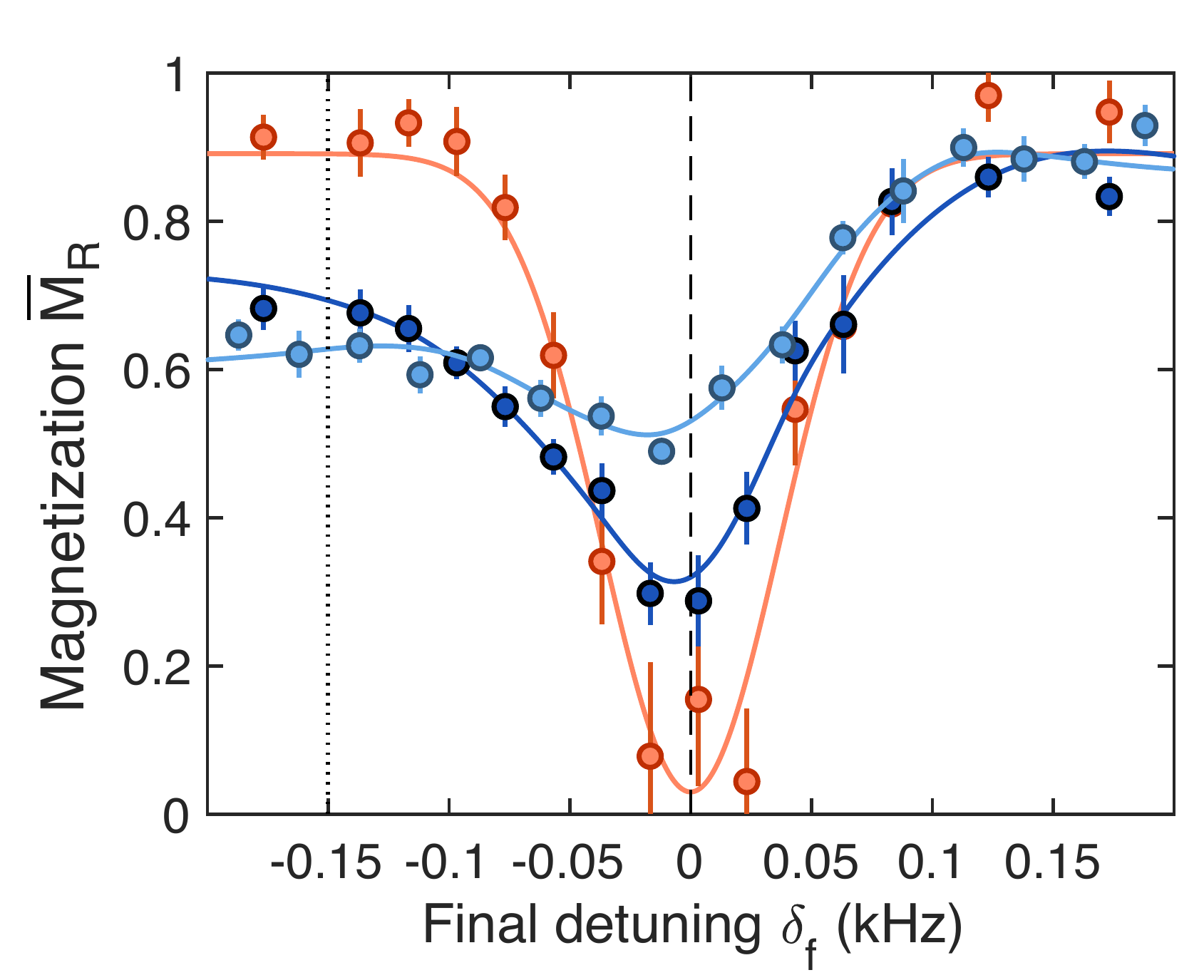}
    \caption{Effects of single atoms on $\overline{M}_R$. Isolated atoms at $35\,E_R$ (orange points) or coupled atoms at $11\,E_R$ with lower (dark blue) or higher (light blue points) initial condensate fraction. The solid lines are phenomenological fits to guide to the eye. The dashed vertical line is the $\langle S_z\rangle=0$ point for isolated sites ($35\,E_R$ lattice) and the dotted line is the $\langle S_z\rangle=0$ point for coupled sites in 1D chains ($11\,E_R$ lattice). }
    \label{fig:singlons1}
\end{figure}

This is shown in Fig.\ref{fig:singlons1}. For deep lattices (orange), the return magnetization at zero detuning (dashed line) is 0. At shallower lattices, the dip of the return magnetization at zero detuning signals the presence of isolated particles. By varying the preparation protocol, the number of single particles can be increased. The Mott insulator is created by loading a Bose-Einstein condensate into an optical lattice. We can vary the condensate fraction, thus increasing the thermal atoms and holes in the Mott insulator. We can estimate the fraction of isolated atoms by using Eq.(\ref{eq:mr_delta}) and subtracting a fraction of a fit of the $35\,E_R$ data. Note that the width of the $35\,E_R$ data is limited by magnetic field noise, estimated to $3.5\,$mG rms. For lower initial condensate fraction (dark blue points), this gives us a single-atom fraction of approximately $30\%$. For higher initial condensate fraction (dark blue points), this gives us a single-atom fraction of approximately $8-10\%$. Therefore, the measured $\overline{M}_R$ could be too high by at most $10\%$. Improved detection methods, such as a quantum gas microscope, could give a better picture.

\section{Noise sources }\label{appG}
\begin{figure}
\centering
\begin{overpic}[width=0.5\textwidth]{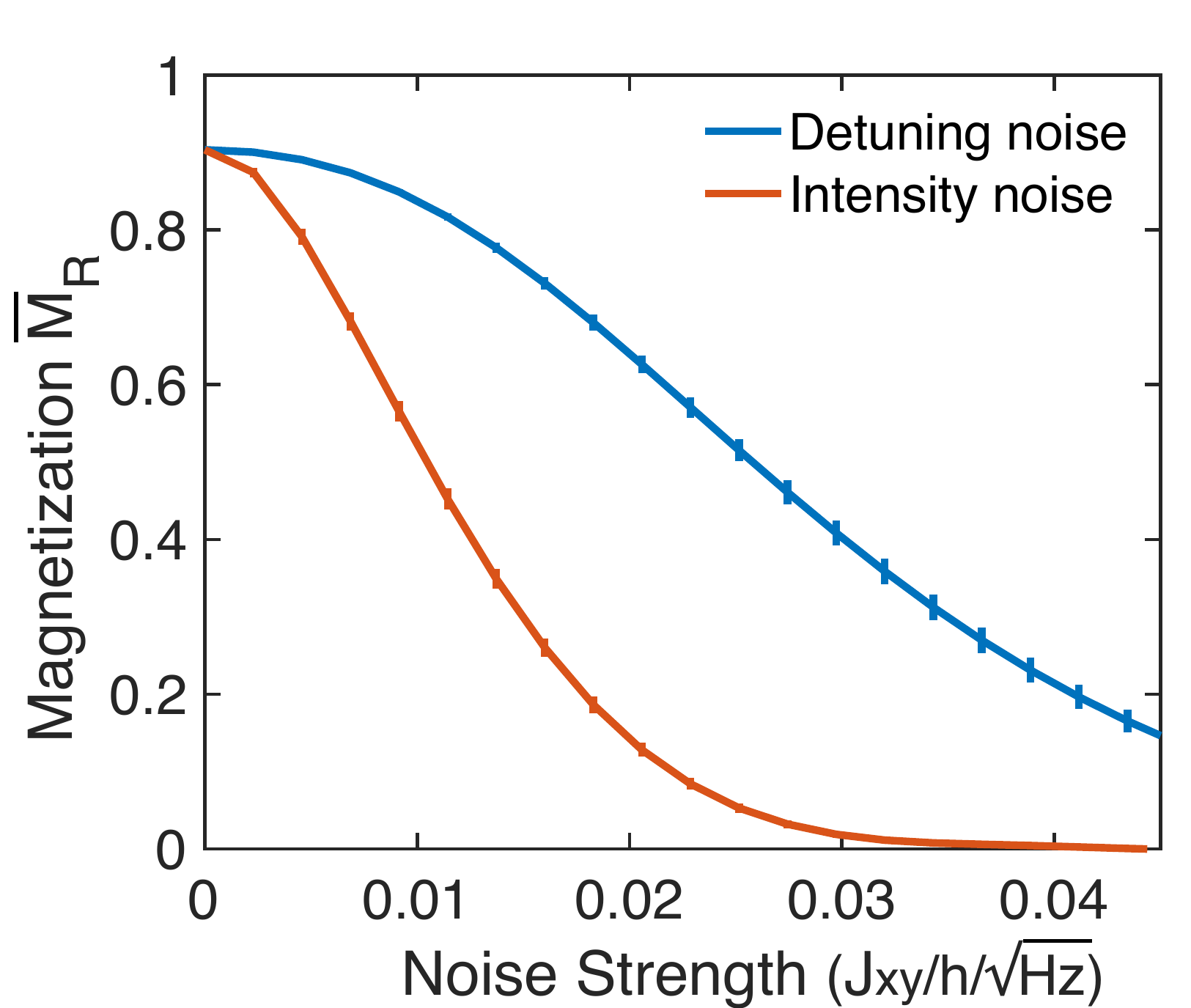}
\end{overpic}
\caption{Noise sources. Plotted are the calculated effects of detuning noise (x-field noise) and intensity noise of the microwave field (x-field noise) on the return magnetization $\overline{M}_R$, assuming white noise and optimized exponential ramps. Our observed $\overline{M}_R$ around 0.5 are probably limited by intensity noise.}
\label{fig:noise_sources}
\end{figure}

The main sources of noise in our adiabatic protocol are detuning and intensity noise of the microwave field, which map to noise in the $z-$ and $x-$field respectively. For individual spins, we expect noise in the $z-$field to be dominant. This is evident in the dephasing of individual spins rotated to the $xy-$plane, as shown in Fig.2(c) in the main text. For coupled spins, the adiabatic preparation protocol relies on the noise being smaller than the gap to the next excited state, which is the smallest at the final stages of the ramp in our case. Numerical simulations show the effects of detuning and intensity noise on the return magnetization $\overline{M}_R$ given our preparation protocol with optimized ramp times and assuming white noise, Fig. \ref{fig:noise_sources}. For the same power spectral density, intensity noise results in a larger decrease of $\overline{M}_R$.

We estimate the power spectral density of the detuning noise by measuring the current in the coils creating the magnetic field. We see a flat profile up to several kHz, beyond which, due to the large inductance of the coils, fluctuations are suppressed. We estimate that the rms-noise of 3.5 mG corresponds to power spectral density of $0.002\, J_{xy}/h/\sqrt{\rm Hz}$, which is not limiting. 

To estimate the amount of intensity noise, we measure the decay time of Rabi oscillations of individual atoms. For Rabi oscillations in a two-level system, we can include intensity noise in the Rabi frequency:
\begin{equation}
    \overline{\Omega(t)} = \Omega(t) + \Delta \Omega \epsilon(t) 
\end{equation}
where $\overline{\epsilon(t) \epsilon(t')} = S_0 \delta(t-t')$, which describes white noise with strength  $S_0$. For this type of noise, the Rabi oscillations envelope will decay as $e^{-t/\tau}$, where $\tau$ is the characteristic decay time. We can then express the decay time as:
\begin{equation}
    \tau = \frac{2}{S_0(\Delta \Omega)^2}
\end{equation}
and the noise strength as:
\begin{equation}
   f_N \equiv \sqrt{\frac{2}{\tau}} \frac{1}{J_{xy}/\hbar }
\end{equation}
in units of $J_{xy}/h\sqrt{\rm Hz}$ where $J_{xy}/\hbar = 2\pi$ 84.5 Hz. In our experiment, the decay time increases with decreasing Rabi frequency, probably due to decreasing signal-to-noise ratio given by the constant noise added by the power amplifier.  We can put an upper bound on the intensity noise by assuming that the decay of the Rabi oscillations is only due to intensity fluctuations. Assuming white noise, we estimate that the intensity noise is $f_N= $0.0094$\, J_{xy}/h/\sqrt{\rm Hz}$ for large Rabi frequencies and increases to $0.019\, J_{xy}/h/\sqrt{\rm Hz}$ at the very final stages of the ramp. Given that for optimized sweeps a return magnetization of $\overline{M}_R \sim 0.5$ corresponds to Rabi frequency intensity noise of $\sim 0.01\, J_{xy}/\sqrt{\rm Hz}$, we can conclude that our experiment is mainly limited by intensity noise in the microwave field at the final stages of the ramp.

\end{document}